\documentclass[aps,prl,twocolumn,superscriptaddress,showpacs]{revtex4}

\usepackage{graphicx}
\usepackage{dcolumn}
\usepackage{bm}
\usepackage{color}

\begin{document}
\preprint{APS/123-QED}

\title{Temperature-Dependence of Magnetically-Active Charge Excitations in Magnetite across the Verwey Transition}

\author{M. Taguchi}

\email[Corresponding author: ]{mtaguchi@ms.naist.jp}

\affiliation{Material Science, Nara Institute of Science and Technology (NAIST), Ikoma, Nara, 630-0192, Japan}

\affiliation{RIKEN SPring-8 Center, Sayo, Sayo, Hyogo 679-5148, Japan}

\author{A. Chainani}

\affiliation{RIKEN SPring-8 Center, Sayo, Sayo, Hyogo 679-5148, Japan}

\author{S. Ueda}

\affiliation{National Institute for Materials Science (NIMS), SPring-8, Sayo, Hyogo 679-5148, Japan}

\author{M. Matsunami}

\affiliation{ Institute for Solid State Physics, University of Tokyo, Kashiwa, Chiba 277-8581, Japan}

\author{Y. Ishida}

\affiliation{ Institute for Solid State Physics, University of Tokyo, Kashiwa, Chiba 277-8581, Japan}

\author{R. Eguchi}

\affiliation{RIKEN SPring-8 Center, Sayo, Sayo, Hyogo 679-5148, Japan}

\author{S. Tsuda}

\affiliation{National Institute for Materials Science (NIMS), Tsukuba, Ibaraki, 305-0003, Japan}

\author{Y. Takata}

\affiliation{RIKEN SPring-8 Center, Sayo, Sayo, Hyogo 679-5148, Japan}

\author{M. Yabashi}

\affiliation{RIKEN SPring-8 Center, Sayo, Sayo, Hyogo 679-5148, Japan}

\author{K. Tamasaku}

\affiliation{RIKEN SPring-8 Center, Sayo, Sayo, Hyogo 679-5148, Japan}

\author{Y. Nishino}

\affiliation{RIKEN SPring-8 Center, Sayo, Sayo, Hyogo 679-5148, Japan}

\author{T. Ishikawa}

\affiliation{RIKEN SPring-8 Center, Sayo, Sayo, Hyogo 679-5148, Japan}

\author{H. Daimon}

\affiliation{Material Science, Nara Institute of Science and Technology (NAIST), Ikoma, Nara, 630-0192, Japan}

\author{S. Todo}

\affiliation{ Institute for Solid State Physics, University of Tokyo, Kashiwa, Chiba 277-8581, Japan}

\author{H. Tanaka}

\affiliation{ISIR-Sanken, Osaka University, 8-1 Mihogaoka, Ibaraki, Osaka, 567-0047, Japan}

\author{M. Oura}

\affiliation{RIKEN SPring-8 Center, Sayo, Sayo, Hyogo 679-5148, Japan}

\author{Y. Senba}

\affiliation{JASRI/SPring-8, Sayo, Sayo, Hyogo 679-5198, Japan}

\author{H. Ohashi}

\affiliation{JASRI/SPring-8, Sayo, Sayo, Hyogo 679-5198, Japan}

\author{S. Shin}

\affiliation{RIKEN SPring-8 Center, Sayo, Sayo, Hyogo 679-5148, Japan}
\affiliation{ Institute for Solid State Physics, University of Tokyo, Kashiwa, Chiba 277-8581, Japan}

\date{\today} 

\begin{abstract}
We studied the electronic structure of bulk single crystals and epitaxial films of Fe$_3$O$_4$. Fe $2p$ core level spectra show clear differences between hard x-ray (HAX-) and soft x-ray (SX-) photoemission spectroscopy (PES). The bulk-sensitive spectra exhibit temperature ($T$)-dependence across the Verwey transition, which is missing in the surface-sensitive spectra. By using an extended impurity Anderson full-multiplet model, and in contrast to earlier peak assignment, we show that the two distinct Fe-species (A- and B-site) and the charge modulation at B-site are responsible for the newly found double peaks in the main peak above $T_V$ and its $T$-dependent evolution. The Fe $2p$ HAXPES spectra show a clear magnetic circular dichroism (MCD) in the metallic phase of magnetized 100-nm-thick films. The model calculations also reproduce the MCD and identify the contributions from magnetically distinct A- and B-sites. Valence band HAXPES shows finite density of states at $E_F$ for the polaronic half-metal with remnant order above $T_V$, and a clear gap formation below $T_V$.  The results indicate that the Verwey transition is driven by changes in the strongly correlated and magnetically active B-site electronic states, consistent with resistivity  and optical spectra.

\end{abstract}


\maketitle
In spite of an extensive research of nearly 80 years,\cite{deB37,pei37} the microscopic description of the metal-insulator (MI, strictly speaking a half-metal to insulator) transition in magnetite (Fe$_3$O$_4$) remains one of the most important open problems in strongly correlated electron systems. Magnetite is a classic mixed-valence oxide which undergoes a first order MI phase transition on cooling below $T_V$$\sim$120 K, at which the electrical resistivity abruptly increases by 2 orders of magnitude.\cite{ver39} Simultaneously, the inverse spinel crystal structure changes from cubic to monoclinic symmetry. Verwey $et$ $al.$\cite{ver47} interpreted this MI transition as a charge ordering of the Fe$^{2+}$ and Fe$^{3+}$ states on the B-site.  However, the original Verwey model was later refuted by several experiments\cite{gar04,wal02}. Recent structural studies\cite{wri02,bla11,sen12} of high resolution neutron and x-ray powder diffraction data have indicated a small charge disproportionation of only $\sim$0.3$e$ between B-site Fe cations with 2+ and 3+ formal valency. 

On the theoretical front, recent studies have emphasized the role of orbital order, multiferroicity and half-metal accompanying the charge-ordering Verwey transition.\cite{jen04,bri08} And while the role of on-site Coulomb correlations as well as long-range Coulomb correlations have been invoked, their relations with the experimental electronic structure are still not conclusive. Thus, despite intensive investigations, fundamental questions about the Verwey transition such as the origin of the charge-orbital ordering, the role of disorder and residual entropy,\cite{sen12,she91} the precise mechanism of  electronic charge transport and  the accurate microscopic interaction driving the MI transition are still debated. While enormous progresses were achieved in resolving the low-$T$ monoclinic structure\cite{iiz82,bla11,wri02}, the superstructure associated with the charge ordering was identified only recently as a $trimeron$ consisting of $t_{2g}$ minority spin electrons delocalized over three B-Fe sites: a donor Fe$^{2+}$ and two neighboring acceptor Fe$^{3+}$  in the formal valency picture.\cite{sen12} Further, using time-resolved soft x-ray diffraction and optical reflectivity, it was shown that the trimerons become mobile across the MI transition in two steps process.\cite{jon13}

 From the spectroscopic viewpoint, two major fundamental questions about the electronic states remain unresolved. 
The first question concerns the peak assignment of the Fe $2p$ core level SX- photoemission spectroscopy (PES)\cite{fuj99,che04} and x-ray magnetic circular dichroism (XMCD) in Fe $L_{2,3}$ absorption edge.\cite{kup97} It is well known that Fe $2p$ SX-PES and XMCD spectra of Fe$_3$O$_4$ exhibit typical spectral shapes, which are frequently used as fingerprints of magnetite. These spectra have been often interpreted in terms of three distinct clusters (A-Fe$^{3+}$, B-Fe$^{2+}$, B-Fe$^{3+}$), with the total spectrum being a sum of three distinct spectra with the stoichiometric ratio 1:1:1 for the different Fe sites. However, this model has been applied even to the high-$T$ phase, where all B-site Fe atoms  are crystallographically equivalent.  Recent resonant x-ray scattering measurements, however, have questioned this ionic model.\cite{gar00,gar04} Thus, all B-site Fe atoms should be considered equivalent above $T_V$. 
Another controversial question concerns the electronic nature of the high-$T$ phase. In a valence band (VB) PES study, it was found a gap of 70 meV opening below $T_V$ in the occupied density of states.\cite{cha95} This matches nicely with half the total gap of 140 meV observed below $T_V$ in optical spectroscopy.\cite{par98} The authors also found a finite spectral weight at $E_F$ above $T_V$ and concluded on a half-metallic state with remnant order.\cite{cha95, wan13} In contrast, Park $et$ $al.$ reported a change in the gap just below and above $T_V$, but the gap was still finite above $T_V$.\cite{par97} The authors accordingly concluded that the Verwey transition  was an insulator to semiconductor phase transition. Recent PES studies also suggested that the Verwey transition is a small polaron insulator-semiconductor phase  transition,\cite{sch05,kim10} although clear phonon side band features have never been observed experimentally. This has resulted into two different interpretations of the Verwey transition as an ``insulator to half-metal" or ``insulator to semiconductor'' transition. 

In this study, we have critically investigated the Fe $2p$ core level and VB electronic states near $E_F$ and answer the two fundamental issues described above. By employing bulk-sensitive HAXPES, we show  that Fe $2p$ core level spectra of Fe$_3$O$_4$  exhibit an new bulk character electronic state which shows $T$-dependence, but is strongly suppressed in surface-sensitive spectra. The new feature is magnetically active  as seen in MCD-HAXPES measured from epitaxial Fe$_3$O$_4$ magnetized films. Comparison with the calculated MCD shows that it originates in the B-site  $3d$ electrons associated with the valence disproportions. We also observed a weak spectral weight transfer but over a large energy scale($\sim$2 eV $>$ 10 times the gap energy of 90 meV) and a finite density of states at $E_F$ in the high-$T$ phase, indicative of strong correlations and a metallic state, respectively.

\begin{figure}
\includegraphics[scale=.39]{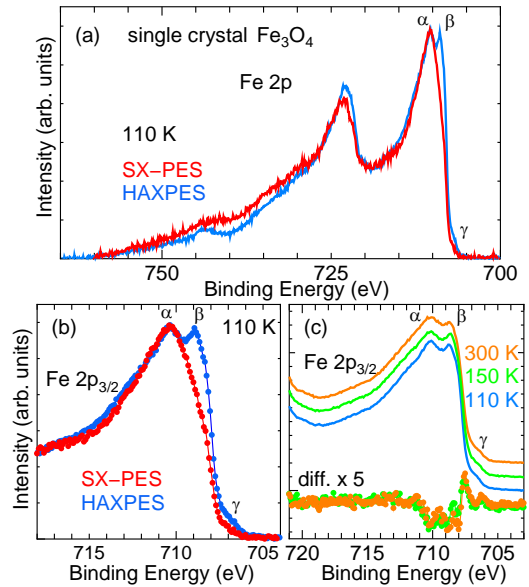}
\caption{\label{fig1}   
(Color online)  Measured  Fe $2p$ PES spectra of single crystal Fe$_3$O$_4$. (a) Comparison between Fe $2p$ HAXPES and SX-PES. The spectra were corrected by subtracting a Shirley-type background. (b) Enlargements of the energy range corresponding to Fe $2p_{2/3}$ line. (c) $T$-dependent Fe $2p$ HAXPES spectra across the Verwey transition. Bottom panel: Difference of 300 K and 150 K spectra to the 110 K spectrum.  }
\end{figure}

High quality single crystals of magnetite were synthesized by the floating zone method and the stoichiometry is Fe$_3$O$_{4.002}$.\cite{tod95,tod01,inf}
The 100-nm-thick Fe$_3$O$_4$ films were fabricated by a pulsed laser deposition technique on a MgO(100) substrate.\cite{ish05,tak06,tak07} HAXPES measurements for the single crystal Fe$_3$O$_4$ were performed 
in a vacuum of 1$\times$10$^{-10}$ Torr
at undulator beam line BL29XU, SPring-8 using a Scienta R4000-10kV electron analyzer.\cite{ish05i,tak05i} The total energy resolution, $\Delta$$E$ was set to $\sim$170 meV. 
SX-PES was performed at BL17SU, with $\Delta E \sim$200 meV. All PES measurements were done under normal emission geometry to maximize the depth sensitivity. A single crystal of Fe$_3$O$_4$ was fractured $in$ $situ$ and
sample temperature was controlled to $\pm 2 $K during measurements. The MCD-HAXPES experiments for 100-nm-thick films were performed at the undulator beamline BL15XU\cite{ued13} of SPring-8 using the circularly polarized x-rays from the helical undulator. Photon energy was set to 5.95 keV,\cite{inf} The MCD-HAXPES measurements of Fe$_3$O$_4$ in a remanent state were performed at room temperature.  Total energy resolution was set to 240 meV. 

First, we present experimental Fe $2p$ HAXPES and SX-PES spectra of single crystal Fe$_3$O$_4$ at 110 K  obtained with photon energies of $h\nu$=7.94 keV and 1.0 keV in Fig.~1(a) and (b). The kinetic energy of the Fe $2p$ level of SX-PES and HAXPES correspond to a probing depth of  $\sim$7 and $\sim$80 \AA, respectively.
The core level spectra were normalized at the feature ``$\alpha$".  The  SX-PES spectrum agrees well with reported results\cite{fuj99,che04} and shows a weak shoulder at 708.5 eV, which is conventionally assigned to the Fe$^{2+}$ ions.  
 In comparison, the spectral weight of the lower binding energy peak labeled ``$\beta$" in the main line is found to be  strongly enhanced in the bulk sensitive HAXPES spectrum. Its energy position is the same as that of the weak shoulder observed in SX-PES. Furthermore, the presence of the broad low binding energy satellite labeled ``$\gamma$" at 707 eV is also clearly seen. The strong enhancement of the $\beta$ feature and observation of the $\gamma$ feature indicate the importance of the HAXPES measurements. The $\beta$ and $\gamma$ features  show a weak but observable $T$-dependence across $T_V$ as shown in Fig.~1(c). These new features cannot be interpreted in the usual cluster model with three distinct species for the cation (A-Fe$^{3+}$, B-Fe$^{2+}$, B-Fe$^{3+}$) because the observed enhancement of the feature $\beta$ indicates the spectral weight contribution of the B-Fe$^{2+}$ must be increased by a factor of two, if the conventional cluster model assignment is true.

\begin{figure}
\includegraphics[scale=.37]{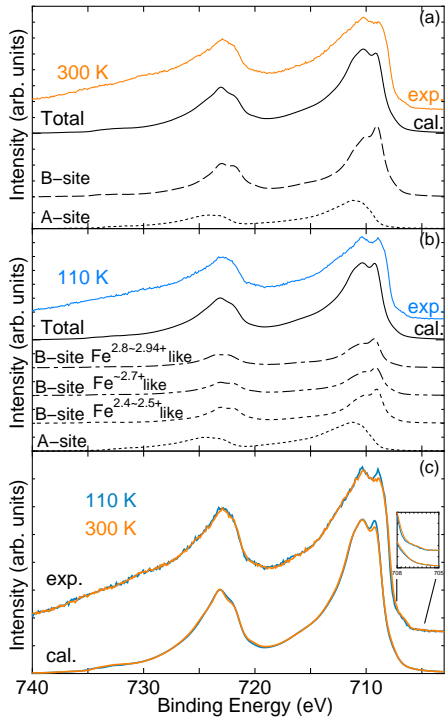}
\caption{\label{fig2}     
(Color online) Calculated $T$-dependent Fe $2p$ spectra of Fe$_3$O$_4$, together with the experimental HAXPES spectra. }
\end{figure}

In order to address the origin of the observed behavior, we used an impurity Anderson model (AIM) with full multiplets  and included a coherent state to describe the spectra. This model is well-established and was successfully used to study the HAXPES spectra for various materials.\cite{hor04,tagprb05,tagprl05,tag08,egu08,tag10,tag13} Details of the model have been described in previous works.\cite{tagprb05,tagprl05,tag08,tag13}
Let us first consider the high-$T$ phase, where all $B$-sites are equivalent. Thus 
 in the calculation, tetrahedral FeO$_4$, and octahedral FeO$_6$ clusters with trigonal distortion  corresponding to the $A$-site and $B$-site were used, respectively. 
 In both A-site FeO$_4$($T_d$) and B-site FeO$_6$($D_{3d}$) clusters, we used, as basis states, six configurations: $3d^5$, $3d^6\underline{L}$, $3d^7\underline{L}^2$, $3d^6\underline{C}$, $3d^7\underline{C}^2$, and $3d^7\underline{CL}$. The $3d^6\underline{C}$ represents the charge transfer between Fe $3d$ and the coherent state on the top of valence band, labeled $C$. An effective coupling parameter $V^*$, for describing the interaction strength between the Fe $3d$ and coherent state is introduced, analogous to the Fe $3d$-O $2p$ hybridization $V$. The parameter values for both A- and B-site clusters are summarized in the supplementary text\cite{inf}. The total spectrum was obtained by making a superposition of the spectra from the above two clusters in the ratio 1:2, because the composition ratio of A-site:B-site is 1:2. A relative energy shift between A- and B-site of 1.4 eV was required, reflecting the known difference of Madelung potential.\cite{ver47} Note that, in the present case, all the B-site Fe cations are treated as equivalent for the metal phase.

The calculated spectrum above $T_V$ is shown in Fig.~2(a) together with the experimental 300 K spectrum. 
 To clarify the peak assignment, the A-site and B-site components are also shown in Fig~2(a). The agreement is remarkable and the experimental features are reproduced well by the calculation over the whole energy range. The sharp peak ``$\beta$" at lower binding energy originates from the B-site Fe atoms and the feature $\alpha$ is a mixture of A-site and B-site components. The screening effect from the coherent state  near $E_F$ leads to the formation of the low energy weak and broad satellite ``$\gamma$".

Next, we consider the low-$T$ insulating phase spectrum.  
 Since the precise XRD studies have suggested the three different type of valences,\cite{bla11,sen12} three kinds of parameter sets  ($i.e.$ Fe$^{2.4\sim2.5+}$-like, Fe$^{\sim2.7+}$-like and Fe$^{2.8\sim2.94+}$-like) were used for the B-sites.\cite{inf}
As we show later, the spectral change of the intensity near $E_F$ is very small. Thus we needed to use finite values of $V^*$ even for the insulating phase, indicating the remnant of the coherent state near $E_F$ in insulating phase. The total calculated spectrum is obtained by a linear combination of the four component spectra with a relative weight of Fe(A):Fe(B$^{2.4\sim2.5}${\small{-like}}):Fe(B$^{\sim2.7}${\small{-like}}):Fe(B$^{2.8\sim2.94}${\small{-like}})=1:$\frac{14}{16}$:$\frac{6}{16}$:$\frac{12}{16}$, respectively [Fig.~2(b)]. Again, the agreement with experiment is remarkable and our calculations nicely reproduce the observed $T$-dependence as shown in Fig.~2(c).

\begin{figure}. 
\includegraphics[scale=.46]{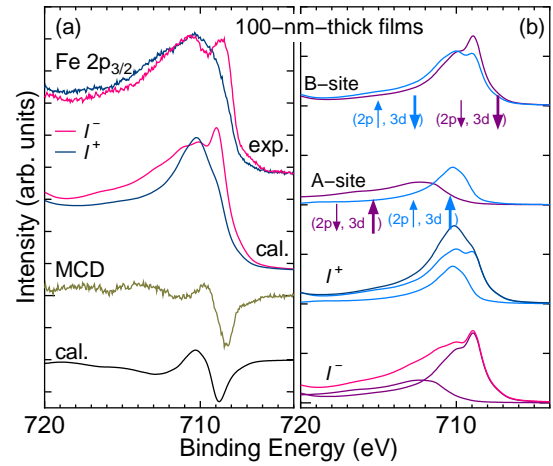}
\caption{\label{fig3}     
(Color online) (a) Fe $2p$ HAXPES spectra of 100-nm-thick films of Fe$_3$O$_4$ for the circularly polarized light. (b) Polarization components of A- and B-site are displayed. }
\end{figure}

Further support to our interpretation can be obtained from independent MCD-HAXPES experiments of 100-nm-thick Fe$_3$O$_4$ films using circularly polarized light, which enable us to obtain clear evidence of the B-site contribution in the main line. Figure~3(a) shows Fe $2p$ HAXPES spectra  of 100-nm-thick film Fe$_3$O$_4$ (probing depth $\sim$60 \AA). The Fe $2p$ spectra were measured for the magnetization parallel and antiparallel to the photon helicity, where $+/-$ signs refer to antiparallel/parallel alignments.  Figure~3(a) clearly shows that the  HAXPES $I^-$ spectra is very similar to the single crystal HAXPES spectra in Fig.~1 and the feature $\beta$  and $\gamma$ exhibit strong MCD. 
The observed MCD behavior is well-reproduced by calculated MCD signal using the same parameter set as for the metallic phase. 
Since the B-site magnetic moment is aligned antiparallel to the A-site moment, the observed MCD allows a complete characterization of the magnetically active states by comparing with the calculated spectral components.  
In Fig.~3(b),  we show the polarization components of A- and B-site spectra. As we mentioned before, the main peak of Fe $2p_{3/2}$ HAXPES spectrum consists of A- and B-site spectra, in which each spectrum has  a spin polarization due to an exchange interaction between $2p$ core hole spin and $3d$ spins. In the  B-site spectrum, the lower binding energy peak at 708.5 eV is mainly composed by the parallel spin component of (Fe$2p$, Fe$3d$$)$$=$$($$\downarrow$, {\Large$\downarrow$}) as seen in the top panel of Fig.~3(b). For the A-site spectrum also, the lower binding energy peak is composed of the parallel spin components ($\uparrow$, {\Large$\uparrow$}) (see second panel of Fig.~3(b)).
The resultant  $I^+$ spectrum of A-site ($\uparrow$, {\Large$\uparrow$}) and B-site ($\uparrow$, {\Large$\downarrow$}) shows a main peak at 710 eV and a shoulder at 708.5 eV (see third panel of Fig.~3(b)). On the other hand, the resultant $I^-$ spectrum of  A-site ($\downarrow$, {\Large$\uparrow$})  and B-site ($\downarrow$, {\Large$\downarrow$}) has a double peak structure seen in the bottom panel of Fig.~3(b).
It should be further noted that the HAXPES spectra of thinner 10-nm-thick films did not show a clear low binding energy feature $\beta$ in the previous report,\cite{ued08} but a very weak MCD signal consistent with the present study. This indicates the importance of HAXPES for bulk-sensitive measurements  as well thickness dependence for the intrinsic electronic structure of Fe$_3$O$_4$ and resolves the debate regarding the core level assignments.  
In retrospect, it is clear that the controversies discussed in the literatures concerning the interpretation of the core level SX-PES, x-ray absorption and XMCD stems from these issues.  

\begin{figure}
\includegraphics[scale=.42]{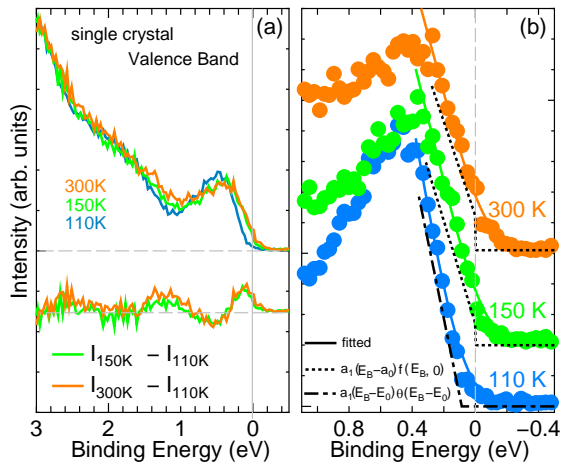}
\caption{\label{fig3}     
(Color online) (a) Measured $T$-dependence of the VB spectra of single crystal Fe$_3$O$_4$ across the Verwey transition, using hard x-ray ($h\nu$=7.94 keV). (b) Solid, dashed, and dotted-dashed curces represent the fitted curves, $(a_1E_B +a_0) f (E_B, 0)$, and $a_0(E_B-E_0)\theta (E_B-E_0)$, respectively.} \end{figure}

In the following, we will focus on the  VB energy region with HAXPES. 
The $T$-dependent VB HAXPES spectra near $E_F$ of single crystal Fe$_3$O$_4$ are shown in Fig.~4(a). The VB consists of the dominantly Fe $3d$ band  with two features;  the minority spin $t_{2g}$ band near $E_F$ (0$-$1.2 eV binding energy)  and a broader feature between 1.2$-$2.5 eV binding energies. The broad O $2p$$-$Fe $4s$ bands at still higher energies are not shown, but we have measured it to be consistent with earlier work.\cite{kim10} The 110 K spectrum shows negligible intensity at $E_F$, indicative of an energy gap of 90 meV.  On increasing temperature across $T_V$, we see a clear spectral weight transfer from high to lower binding energies within the minority spin $t_{2g}$ band, indicating a finite DOS at $E_F$ at $T$=150 K, and a slight increase in DOS at $T$=300 K. The finite DOS at $E_F$ could be simulated using a superposition of a simple power law DOS with a constant\cite{inf,kob02}. We confirmed the slight increase at $E_F$ between $T$=150 K and 300 K. The $T$-dependence within the minority spin $t_{2g}$ half-metallic band is also consistent with the changes seen in the trimerons using XRD, including the changes seen above $T_V$. The $T$-dependent spectra also show weak spectral changes up to an energy scale of 2.5 eV binding energy, which is more than 10 times the gap energy and suggest the importance of strong electron correlations. The results are consistent with optical spectra which indicate a weak Drude peak in the half-metallic phase due to free carriers, albeit indicative of diffusive dynamics like a polaronic metal, as well as changes over large energy scales $\sim$2.5$-$3.0 eV.\cite{par98,ebad13} Furthermore, asymmetry due to electro-hole pair shake-up (the Doniach-\u{S}unji\'c line shape) also clearly increased in the O $1s$ PES for high-$T$ phase, indicating the metallic electronic state.\cite{inf}

In conclusion, HAXPES is used to show that Fe$_3$O$_4$ exhibits new bulk character electronic states, which are strongly suppressed within $\sim$10-nm-thick surface. In order to explain the  observed $T$-dependence and the MCD signal, we need to go beyond the conventional interpretation. By using the extended AIM, we can consistently explain these features in the Fe $2p$ HAXPES spectra. The $T$-dependence originates in the strongly correlated and magnetically active B-Fe $3d$ electronic states.  The VB spectra show a gap formation at low-$T$ and spectral weight transfer from  high to low binding energy side with finite intensity at $E_F$ in the high-$T$ phase, confirming the half-metallic minority spin $t_{2g}$  electronic states. 

M. T. and A. C. would like to thank A. Fujimori, L. H. Tjeng and R. Claessen for useful discussions and illuminating comments. This work was partially supported by KAKENHI (25400335, 26105001). The MCD-HAXPES measurements were performed with the approval of NIMS Beamline Station (Proposal No. 2007B4909).


\begin{references}

\bibitem{deB37}
J. H. de Boer and E. J. W. Verwey, Proc. Phys. Soc. London Sec. A  {\bf 49}, 59 (1937).

\bibitem{pei37} 
R. Peierls, Proc. Phys. Soc. London Sec. A {\bf 49}, 72  (1937).

\bibitem{ver39}
E. J. W. Verwey, Nature {\bf144}, 327 (1939).

\bibitem{ver47}
E. J. W. Verwey, P. W. Haayman, and F. C. Romejin, J. Chem. Phys. {\bf15}, 181 (1947).

\bibitem{wal02}
F. Walz, J. Phys.: Condens. Matter {\bf14}, R285 (2002).

\bibitem{gar04} 
J. Garc\'ia and G. Sub\'ias, J. Phys.: Condens. Matter {\bf16}, R145 (2004).

\bibitem{wri02} J. P. Wright, J. P. Attfield, and P. G. Radaelli, Phys. Rev. Lett. {\bf87}, 266401 (2001).

\bibitem{bla11} J. Blasco, J. Garc\'ia and G. Sub\'ias, Phys. Rev. B {\bf83}, 104105 (2011).

\bibitem{sen12} M. S. Senn, J. Wright, and J. P. Attfield, Nature {\bf 481}, 173 (2012).

\bibitem{jen04}
Horng-Tay Jeng, G. Y. Guo, and D. J. Huang, Phys. Rev. Lett. {\bf93}, 156403 (2004).

\bibitem{bri08}
J. van den Brink and D. I. Khomskii, J. Phys.: Condens. Matter {\bf20}, 434217 (2008). 

\bibitem{she91}
J. P. Shepherd, J. W. Koenitzer, R. Arag\'on, J. Spa\l ek, and J. M. Honig, Phys. Rev. B {\bf43}, 8461 (1991). 

\bibitem{iiz82}
M. Iizumi, T. F. Koetzle, G. Shirane, S. Chikazumi, M. Matsui, and S. Todo, Acta Crystallogr. {\bf38}, 2121 (1982). 

\bibitem{jon13}
S. de Jong, R. Kukreja, C. Trabant, N. Pontius, C. F. Chang, T. Kachel,	 M. Beye, F. Sorgenfrei, C. H. Back, B. Br\"{a}uer, W. F. Schlotter, J. J. Turner, O. Krupin, M. Doehler, D. Zhu, M. A. Hossain, A. O. Scherz,	 D. Fausti, F. Novelli, M. Esposito, W. S. Lee, Y. D. Chuang, D. H. Lu, R. G. Moore, M. Yi, M. Trigo,	 P. Kirchmann, L. Pathey, M. S. Golden, M. Buchholz, P. Metcalf, F. Parmigiani,	W. Wurth, A. F\"{o}hlisch,	 C. Sch\"{u}$\ss$ler-Langeheine, and H. A. D\"{u}rr, Nature Materials {\bf12}, 882 (2013).

\bibitem{fuj99} T. Fujii, F. M. F. de Groot, G. A. Sawatzky, F. C. Voogt, T. Hibma  and K. Okada, Phys. Rev. B {\bf 59}, 3195 (1999).

\bibitem{che04} J. Chen, D. J. Huang, A. Tanaka, C. F.Chang, S. C. Chung, W. B. Wu, and C. T. Chen, Phys. Rev. B {\bf 69}, 085107 (2004).

\bibitem{kup97}
P. Kuiper, B. G. Searle, L.-C. Duda, R. M. Wolf, and P. J. van der Zaag, J. Electron Spectrosc. Relat. Phenom., {\bf 86}, 107 (1997).

\bibitem{gar00}
 J. Garc\'ia, G. Sub\'ias, M. G. Proietti, H. Renevier,Y. Joly, J. L. Hodeau, J. Blasco, M. C. S\'anchez, and J. F. B\'erar, Phys. Rev. Lett. {\bf85}, 578 (2000).

\bibitem{cha95}
A. Chainani, T. Yokoya, T. Morimoto, T. Takahashi, and S. Todo, Phys. Rev. B {\bf 51}, 17976 (1995).

\bibitem{par98} S. K. Park, T. Ishikawa, and Y. Tokura, Phys. Rev. B {\bf 58}, 3717 (1998).

\bibitem{wan13}
W. Wang, J.-M. Mariot, M. C. Richter, O. Heckmann, W. Ndiaye, P. De Padova, A. Taleb-Ibrahimi, P. Le F\`evre, F. Bertran, F. Bondino, E. Magnano, J. Krempask\'y, P. Blaha, C. Cacho, F. Parmigiani, and K. Hricovini, Phys. Rev. B {\bf 87}, 085118 (2013).

\bibitem{par97}
J.-H. Park, L. H. Tjeng, J. W. Allen, P. Metcalf, and  C. T. Chen, Phys. Rev. B {\bf55}, 12813 (1997).


\bibitem{sch05} D. Schrupp, M. Sing, M. Tsunekawa, H. Fujiwara, S. Kasai, A. Sekiyama, S. Suga, T. Muro, V. A. M. Brabers, and R. Claessen, Europhys. Lett. {\bf 70}, 789 (2005).


\bibitem{kim10}
M. Kimura, H. Fujiwara, A. Sekiyama, J. Yamaguchi, K. Kishimoto, H Sugihara, G. Funabashi, S. Imada, S. Iguchi, Y. Tokura, A. Higashiya, M. Yabashi, K. Tamasaku, T. Ishikawa, T. Ito, S. Kimura,  and S. Suga, J. Phys. Soc. Jpn. {\bf 79}, 064710  (2010).

\bibitem{tod95}
S. Todo, K. Siratori, S. Kimura, J. Phys. Soc. Jpn. {\bf 64}, 2118  (1995).

\bibitem{tod01}
S. Todo, N. Takeshita, T. Kanehara, T. Mori, and N. Mori, J. Appl. Phys. {\bf89}, 7347 (2001).

\bibitem{inf}
See Supplemental Material [url], which includes Refs. [2-4], [25] and [42].

\bibitem{ish05}  M. Ishikawa, H. Tanaka, and T. Kawai, Appl. Phys. Lett. {\bf86}, 222504 (2005).

\bibitem{tak06} J. Takaobushi, H. Tanaka, T. Kawai, S. Ueda, J. J. Kim, M. Kobata, E. Ikenaga, M. Yabashi, K. Kobayashi, Y. Nishino, D. Miwa, K. Tamasaku, and T. Ishikawa, Appl. Phys. Lett. {\bf89}, 242507 (2006).

\bibitem{tak07} J. Takaobushi, M. Ishikawa, S. Ueda, E. Ikenaga, J. J. Kim, M. Kobata, Y. Takeda, Y. Saitoh, M. Yabashi, Y. Nishino, D. Miwa, K. Tamasaku, T. Ishikawa, I. Satoh, H. Tanaka, K. Kobayashi, and T. Kawai, Phys. Rev. B {\bf76}, 205108  (2007).

\bibitem{ish05i}
T. Ishikawa, K. Tamasaku, M. Yabashi , Nucl. Instrum. Methods A {\bf547}, 42 (2005).

\bibitem{tak05i} 
Y. Takata M. Yabashi, K. Tamasaku, Y. Nishino, D. Miwa, T. Ishikawa, E. Ikenaga, K. Horiba, S. Shin, M. Arima, K. Shimada, H. Namatame, M. Taniguchi, H. Nohira, T. Hattori, S. Sodergen, B. Wannberg, and K. Kobayashi, Nucl. Instrum. Methods A {\bf547}, 50 (2005).

\bibitem{ued13}
S. Ueda, J. Electron Spectorosc. Rel. Phenom. {\bf190}, 235 (2013).

\bibitem{hor04} K. Horiba, M. Taguchi, A. Chainani, Y. Takata, E. Ikenaga, D. Miwa, Y. Nishino, K. Tamasaku, M. Awaji, A.Takeuchi, M.Yabashi, H. Namatame, M. Taniguchi, H. Kumigashira, M. Oshima, M. Lippmaa, M. Kawasaki, H. Koinuma, K. Kobayashi, T. Ishikawa, and S. Shin, Phys. Rev. Lett. {\bf93}, 236401 (2004).

\bibitem{tagprb05} M. Taguchi, A. Chainani, N. Kamakura, K. Horiba, Y. Takata, M. Yabashi, K. Tamasaku, Y. Nishino, D. Miwa, T. Ishikawa, S. Shin, E. Ikenaga, T. Yokoya, K. Kobayashi, T. Mochiku, K. Hirata, and K. Motoya, Phys. Rev. B {\bf71}, 155102 (2005).

\bibitem{tagprl05} M. Taguchi, A. Chainani, K. Horiba, Y. Takata, M. Yabashi, K. Tamasaku, Y. Nishino, D. Miwa, T. Ishikawa, T. Takeuchi, K. Yamamoto, M. Matsunami, S. Shin, T. Yokoya, E. Ikenaga, K. Kobayashi, T. Mochiku, K. Hirata, J. Hori, K. Ishii, F. Nakamura, and T. Suzuki, Phys. Rev. Lett. {\bf95}, 177002 (2005).

\bibitem{tag08} M. Taguchi, M. Matsunami, Y. Ishida, R. Eguchi, A. Chainani, Y. Takata, M. Yabashi, K. Tamasaku, Y. Nishino, T. Ishikawa, Y. Senba, H. Ohashi, and S. Shin, Phys. Rev. Lett. {\bf100}, 206401 (2008).

\bibitem{egu08} R. Eguchi, M. Taguchi, M. Matsunami, K. Horiba, K. Yamamoto, Y. Ishida, A. Chainani, Y. Takata, M. Yabashi, D. Miwa, Y. Nishino, K. Tamasaku, T. Ishikawa, Y. Senba, H. Ohashi, Y. Muraoka, Z. Hiroi, and S. Shin, Phys. Rev. B {\bf78}, 075115 (2008).

\bibitem{tag10} M. Taguchi, A. Chainani, M. Matsunami, R. Eguchi, Y. Takata, M. Yabashi, K. Tamasaku, Y. Nishino, T. Ishikawa, S. Tsuda, S. Watanabe, C.-T. Chen, Y. Senba, H. Ohashi, K. Fujiwara, Y. Nakamura, H. Takagi, and S. Shin, Phys. Rev. Lett. {\bf104}, 106401 (2010).

\bibitem{tag13} M. Taguchi, Y. Takata and A. Chainani,  J. Electron Spectrosc. Relat. Phenom. {\bf 190}, 242 (2013).

\bibitem{ued08} S. Ueda, H. Tanaka, J. Takaobushi, E. Ikenaga, J. Kim, M. Kobata, T. Kawai, H. Osawa, N. Kawamura, M. Suzuki, K. Kobayashi, Appl. Phys. Express {\bf1}, 077003 (2008). 

\bibitem{kob02} K. Kobayashi, T. Susaki, A. Fujimori, T. Tonogai, and H. Takagi, Europhys. Lett. {bf59}, 868 (2002).

\bibitem{ebad13} J Ebad-Allah, L Baldassarre, M Sing, R Claessen, V A M Brabers and C A Kuntscher, J. Phys: Condens. Matter {\bf25}, 035602 (2013).


\end{references}
\end{document}